\documentstyle[hip-artc]{article}
\volnumber{15}  \edyear{2002} \frompage{000} \topage{000}                %|
\recrevdate{15 July 2001}                                                %|
\def\rightmark{Order generated by random many-body dynamics}
\def\leftmark{A. Volya and V. Zelevinsky}

\title{Order generated by random many-body dynamics}
\authors{\twerm
{Alexander Volya$^1$ and Vladimir Zelevinsky$^{2}$ %
\index{Volya, A.}
\index{Zelevinsky, V.}
}\\[2.812mm]
{\normalsize
\hspace*{-8pt}$^1$ Physics Division, Argonne National Laboratory,
Argonne, Illinois 60439, USA \\[0.2ex]
\hspace*{-8pt}$^2$ National Superconducting Cyclotron Laboratory and
Department of Physics and Astronomy,
Michigan State University,
E-Lansing, Michigan 48824, USA
}}

\abstract{New features related to collective properties generated
in the systems driven by random dynamics are observed and their
implications for further understanding of interplay between
coherence and chaos are discussed. }

\makeindex
\begin{document}

\maketitle

One of the most extensively developing directions in Wigner's
scientific legacy is associated with random matrix theory (RMT).
From a current viewpoint, the RMT corresponds to the limit of
extreme chaoticity in the dynamics of a complicated quantum
system. During the last decade, the center of interest shifted
from the well understood cases, such as quantum billiards or
the hydrogen atom in a magnetic field, where the source of chaoticity
is in the external conditions of incompatible symmetry, to
many-body systems where the interparticle interaction is a driving
force for the onset of quantum chaos. Nuclear physics, which long
ago was the first arena to host studies of chaotic spectra of
complex quantum systems, provides new insights into manifestations
of many-body quantum chaos. Inversely, ideas of chaos give rise to
new tools in understanding nuclear structure. Using  examples
taken from nuclear physics we demonstrate here an amazing
coexistence of quantum chaos and order in a strongly interacting
mesoscopic many-body system.

We consider a nuclear system described by a Hamiltonian with
residual interactions that are defined at random but preserve the
rotational invariance of the system. It turns out that with a high
probability this results in orderly features bearing a striking
resemblance to observed properties of nuclei. We limit our
discussion here to a simple system containing just a single
degenerate $j$-level, thus capable of accommodating $\Omega=2j+1$
fermions. The general two-body Hamiltonian is defined as
\begin{equation}
H=\frac{1}{2}\sum_L V_L \sum_{\Lambda} P_{L \Lambda}^\dagger P_{L
\Lambda}, \quad {\rm where}\,\, P_{L \Lambda}=\left (a_{j}a_{j}
\right )_{L \Lambda}                         \label{hamiltonian}
\end{equation}
is a destruction operator for a pair of nucleons with angular
momentum $L$ and projection $\Lambda.$ The Fermi-statistics
permits only even values $L=0,2,...,2j-1$. The $j+1/2$ parameters
$V_L$ are selected at random with zero average and mean square
deviation defining the unit of energy. The results are not
critically sensitive to the details of the random ensemble, being
mainly determined by underlying symmetries.

\begin{figure}[htb]
\vspace*{-0.8cm}
%                 \insertplot{lalaga.EPS}
\epsfxsize=11.5cm \epsfbox{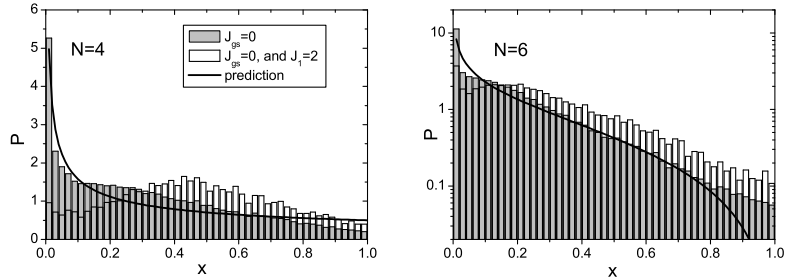} \vspace*{-1.1cm}
\caption[]{Distribution of overlaps $x$ of the actual ground
states of spin $J_{0}=0$ and the fully paired state for the case
$j=15/2$ with $N=4$ (left) and $N=6$ (right) particles. The shaded
histograms include all realizations resulting in $J_0=0\,;$ the
unshaded distribution involves an additional requirement $J_1=2$
for the spin of the first excited state. The solid lines show the
statistically expected curves for the random orientation of a
ground state vector, $P(x)\sim 1/\sqrt{x}$ ($N=4$, dimension
$d=3$) and $P(x)\sim (1-x)^{3/2}/\sqrt{x},$ ($N=6$, $d=6$) in
Hilbert space spanned by $J=0$ states.}           \label{fig1}
\end{figure}
\vspace*{-0.5cm}

It was first noted in Ref.~\cite{johnson98} that in this 
ensemble with an even number of particles the probability to
encounter a ground state with spin $J_0=0$ significantly exceeds
the statistical expectation given by the fraction of $J=0$ states
in the available space. Features of odd and odd-odd (including
isospin) systems are similar if viewed as one or two particles
outside an even core and in all cases close to properties of
actual nuclei governed by a deterministic nuclear Hamiltonian.
These observations attracted a lot of attention; see for example
\cite{xxx} and references therein. The statistical model
\cite{mulhall00} makes a first step in explaining this phenomenon
with the aid of an effective Hamiltonian $\tilde{H}$ for classes
of states with certain values of conserved quantities $N$ and
${\bf J}^{2}$. From the original Hamiltonian (\ref{hamiltonian})
the most coherent parts of interaction, the monopole,
$\tilde{V}_0$, and rotational (effective moment of inertia),
$\tilde{V}_1$, are extracted,
$$
\langle \tilde{H} \rangle_{N J}=\tilde{V}_0 N(N+1) + \tilde{V}_1
{\bf J}^2,\, \tilde{V}_0=\sum_L
\frac{(2L+1)V_L}{\Omega(\Omega-1)},\, \tilde{V}_1=\sum_L
\frac{(2L+1)V_L}{2\Omega^2 {\bf j}^4} ({\bf L}^2-2{\bf j}^2).
$$
Clearly, in this approximation the ground state spin has a 50-50\%
chance to be either $J_0=0$ or take the maximum possible value
$J_0=J_{max},$ depending on the sign of $\tilde{V}_1.$ Numerical
studies as well as other theoretical methods support this
description for $J_{0}=0$. The excess of
$J_{0}=J_{max}$ is also observed but with the probability
much less than 50\%; in a single-$j$ case this state is
constructed uniquely and the mean-field approach has to be
corrected. The whole picture is close to reality but still
incomplete, and a more detailed analysis shows
deviations from those average predictions. In particular, the
low-lying states reveal correlations that go beyond this simple
model.

Pairing correlations in the $L=0$ channel seem to be the first
candidate for a source of the correlations. The kinematics of a
single-$j$ level favor pairing; for example out of the $j+1/2$
linearly independent choices of $V_L$ in (\ref{hamiltonian}) more
than 2/3 conserve the seniority (number of unpaired particles) 
\cite{volya02_EPM}. Pairing,
however, does not explain the $J_0=0$ preponderance in typical
shell-model situations with random interactions. In Fig.
\ref{fig1} the statistics of the overlaps $x=|\langle \psi |\psi_p
\rangle |^2$ is shown, where $|\psi\rangle $ is the wave function
of a $J_0=0$ state and $|\psi_p \rangle $ is the $J=0$ ground
state for the pure pairing Hamiltonian defined by $V_0=-1$ and
$V_{L \ne 0}=0 .$ The amount of pairing in the actual state
$|\psi\rangle$ agrees with the prediction for the random
orientation of the state vector, and thus indicates no preference
for pairing. This randomness coexists with the previously
mentioned seniority rules. In many-level cases 
off-diagonal phase-insensitive pair transfer
processes increase the predominance of $J_{0}=0$. 

 \begin{figure}[htb]
\vspace*{-0.9cm}
%                 \insertplot{lalaga.EPS}
\epsfxsize=13cm \epsfbox{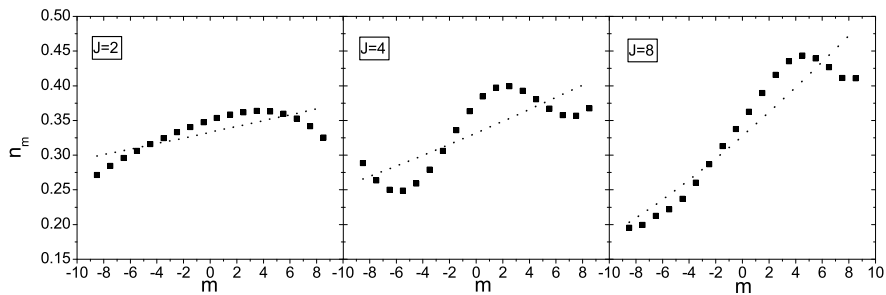} \vspace*{-1.3cm}
\caption[]{Average occupancies $n_m$ in the lowest $J=M$ state for
the system $j=17/2$ and $N=6$. Dotted lines show the statistical
prediction.} \label{fig2}
\end{figure}
\vspace*{-0.5cm}

\begin{figure}[htb]
\vspace*{-0.8cm}
%                 \insertplot{lalaga.EPS}
\epsfxsize=11.5cm \epsfbox{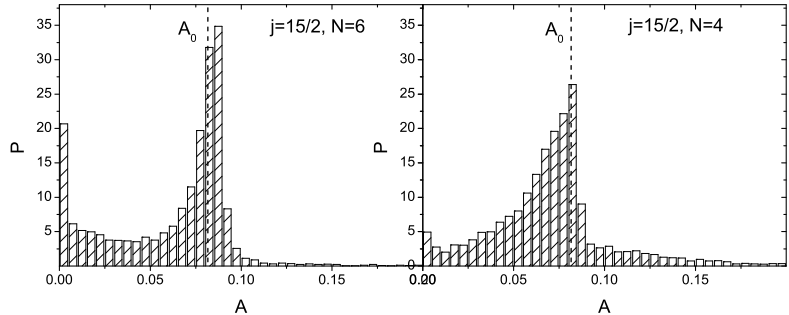} \vspace*{-1.3cm}
\caption[]{The normalized histograms display the distribution of
the ratio $A=Q^2/B({\rm E2})$ for all random realizations with
$J_0=0$ and $J_1=2,$ for the systems of $N=4$ and $N=6$ particles
on a $j=15/2$ orbital. The vertical dashed lines show the Alaga
prediction for a rotor.} \label{fig3}
\end{figure}

The dynamics of fermions in each realization of the ensemble
creates a mean field, which shows substantial coherence beyond
that predicted by the statistical model. Fig. \ref{fig2} compares
average over the ensemble occupation numbers $n_m=\langle
\psi|a^\dagger_{j m} a_{j m}| \psi \rangle$ for the lowest state
$\psi$ of given symmetry $J=M$ with the statistical prediction.
It is clear that these deviations are systematic. This is a
generic result, except for some special cases, such as for a
half-occupied shell, where there are no deviations because of
particle-hole symmetry. For Fig. \ref{fig3}, as well as for the
unshaded histograms in Fig. \ref{fig1}, the realizations that lead
to $J_0=0$ along with $J_1=2$ for the first excited state are
selected. The number of these realizations exceeds the statistical
expectation. For both figures $j=15/2$ is used; for $N=4$ the
fraction of $J_{0}=0$ is 48.4\%, and the combination of $J_0=0$
and $J_1=2$ occurs in 6.7\%; for $N=6$, $J_0=0$ in 68.7\%, while
the probability of the combination $J_0=0$ and $J_1=2$ is 9.2\%.
Fig. \ref{fig3} shows the distribution of the ratio $A=Q^2/B(E2)$
where $Q$ is the quadrupole moment of the $J_1=2$ state and
$B(E2)$ is the strength of the transition from this state to the
ground state. The dashed line indicates $A_0=4/49,$ the rigid
rotor value (Alaga rule). This picture shows a tendency of random
interactions to create a deformed state with rotational structure.
The onset of deformation provides an explanation for an
enhancement of pairing correlations in this class of random
realizations, seen in Fig. \ref{fig1}, as any deformation in such
system leads to enhancement of pairing correlations
\cite{volya02_EPM}. This is a simple geometric effect that scales
as $1/\Omega.$

\vspace*{-0.5cm}

To conclude, in this work we point out new collective features
emerging in finite many-body systems with random dynamics. We
discuss their origin and outline the route for understanding
these phenomena that must go beyond the standard statistical
descriptions and  include non-linear correlations that shape the
many-body mean field. The observed empirical results that come
from randomness but exhibit realistic collective features
emphasize the idea dating back to Wigner that robust realistic
properties are not necessarily a result of a unique nature-chosen
interaction, but rather a generic manifestation of kinematic rules
and symmetries. We need to extend the RMT for the presence of
non-mixed classes of states governed by the same Hamiltonian. This
lays the ground for further mutually beneficial applications of
quantum chaos to strongly interacting many-body systems.\\
{\bf Acknowledgments}\\
A.V. thanks R. Chasman for useful discussions. The calculations
were performed on the parallel computers of the Mathematics and
Computer Science Division, Argonne National Laboratory. This work
is supported by the U. S. Department of Energy, Nuclear Physics
Division, under contract No. W-31-109-ENG-38, and by the National
Science Foundation, grant PHY-0070911.
%\bibliographystyle{h-elsevier3}
%\bibliography{volya,random}

\vspace*{-0.3cm}
\begin{thebibliography}{99}
\vspace*{-0.3cm}
\bibitem{johnson98}
C. Johnson, G. Bertsch and D. Dean,
{\it Phys. Rev. Lett.}  {\bf 80} (1998) 2749.

\bibitem{xxx}
Y.M. Zhao, {\sl et.al.}, {\it Phys. Rev. C}  {\bf 66} (2002)
034302; V. Kota  {\sl et.al.}, {\it Phys. Rev. E} {\bf 65} (2002)
026130; R. Bijker,  {\sl et.al.}, {\it Phys. Rev. C}  {\bf 60}
(1999) 021302(R).

\bibitem{mulhall00}
D. Mulhall, A. Volya and V. Zelevinsky, {\it Phys. Rev. Lett.}
{\bf 82} (2000) 4016.

\bibitem{volya02_EPM}
A. Volya, {\it Phys. Rev. C}  {\bf 65} (2002) 044311.

\end{thebibliography}

\vfill\eject
\end{document}